\documentclass[12pt]{article}

\newcommand{\be}{\begin{equation}}
\newcommand{\ee}{\end{equation}}
\newcommand{\bi}[1]{\vspace{-3mm} \bibitem{#1}}

\usepackage{epsfig,amsmath}
\usepackage{graphics}

\begin{document}

\begin{center}
Celestial Mechanics and Dynamical Astronomy, 
Vol.94. No.1. (2006) pp.1-15.
\end{center}

\begin{center}
{\Large \bf Gravitational Field of \\
Fractal Distribution of Particles}
\vskip 5 mm

{\large \bf Vasily E. Tarasov}\\

\vskip 3mm
{\it Skobeltsyn Institute of Nuclear Physics, \\
Moscow State University, Moscow 119992, Russia } \\
{E-mail: tarasov@theory.sinp.msu.ru}
\end{center}
\vskip 7 mm

\begin{abstract}
In this paper we consider the gravitational field
of fractal distribution of particles.
To describe fractal distribution, we use the fractional integrals.
The fractional integrals are considered as approximations of
integrals on fractals.  
Using the fractional generalization of the Gauss's law, 
we consider the simple examples of the fields of homogeneous
fractal distribution. 
The examples of gravitational moments 
for fractal distribution are considered. 
\end{abstract}

PACS: 98.62.Ck; 05.45.Df;  45.50.-j; 98.80.Jk; 98.65.-r  \\

%05.45.Df Fractals
%45.50.-j Dynamics and kinematics of a particle and a system of particles
%46.05.+b General theory of continuum mechanics of solids

%98.62.Ck Masses and mass distribution 
%98.80.Jk Mathematical and relativistic aspects of cosmology
%02.90.+p Other topics in mathematical methods in physics
%98.65.-r  Large scale structure of the Universe.

Keywords: fractals, fractional integrals, gravitational multipole moment\\

\section{Introduction}

The aim of this paper is to consider the general properties of gravitational 
field that is generated by a fractal distribution of particles. 
This problem is nowadays particularly relevant.
In fact, there is a general agreement that galaxy distribution exhibits
fractal behavior up to a certain scale \cite{slmp98,cp92}. 
The eventual presence of a transition 
scale towards homogeneity and the exact value of the fractal 
dimension  are  still   matters of debate. 
Moreover it has been observed that cold gas clouds of the interstellar 
medium has a fractal structure, with $ 1.5 \le D \le 2$ in a 
large range of length scales \cite{lar,Scalo}. 
Derivatives and integrals of fractional order \cite{SKM} have found 
many applications in recent studies of fractal structures.
Fractional analysis can have numerous applications: 
kinetic theories \cite{Zaslavsky1,Zaslavsky2,Physica,JPA05-2};  
statistical mechanics \cite{chaos,PRE05,JPCS}; 
dynamics in complex media \cite{Nig,PLA05,PLA05-2,AP05,Chaos05}; 
and many others. 
The new type of problem has increased rapidly in areas 
in which the fractal features of a process or the medium 
impose the necessity of using non-traditional tools 
in "regular" smooth physical equations. 
In order to use fractional derivatives and fractional integrals 
for fractal distribution, we must use some 
continuous medium model \cite{PLA05,AP05,Chaos05}. 

We propose to  describe the fractal distribution by a "fractional" 
continuous medium \cite{PLA05}, where all characteristics and 
fields are defined 
everywhere in the volume but they follow some generalized 
equations which are derived by using fractional integrals.
In many problems the real fractal structure of medium
can be disregarded and the fractal distribution can be replaced by
some "fractional" continuous mathematical model.
Smoothing of microscopic characteristics over the
physically infinitesimal volume transforms the initial
fractal distribution into "fractional" continuous model \cite{PLA05}
that uses the fractional integrals.
The order of fractional integral is equal
to the fractal dimension of distribution. 
The fractional integrals allow us to take into account the
fractality of the distribution.
Fractional integrals are considered as an approximation of 
integrals on fractals \cite{RLWM,Nig4}. 
It was proved in \cite{RLWM} that integrals
on net of fractals can be approximated by fractional integrals.
In Ref. \cite{chaos}, we proved that fractional integrals
can be considered as integrals over the space with fractional
dimension up to numerical factor. In order to prove, we use the 
formulas of dimensional regularizations \cite{Col}.

In this paper, we consider gravitational field
of fractal distribution of particles.
Fractal distribution is described by fractional continuous medium model
\cite{PLA05,AP05,Physica,Chaos05}. 
In the general case, the fractal distribution of particles cannot 
be considered as continuous medium.
There are points and domains that are not filled of particles.
In Ref. \cite{PLA05}, we suggest to consider the fractal distributions  
as special ("fractional") continuous media.
We use the procedure of replacement of the distribution 
with fractal mass dimension by some continuous model that 
uses fractional integrals. This procedure is a fractional 
generalization of Christensen approach \cite{Chr}.
Suggested procedure leads to the fractional integration and 
differentiation to describe fractal distribution.

In Section 2, the density of fractal distribution of mass is considered. 
In Section 3, we consider the simple examples of the gravitational 
field of homogeneous fractal distribution. 
In Section 4, the examples of gravitational quadrupole moments
for fractal distribution are considered. 
Finally, a short conclusion is given in Section 5.

\section{Mass and Balance of Mass for Fractal Distribution}

\subsection{Mass of Fractal Distribution}

Let us consider a fractal distribution of particles.
For example, we can assume that  particles 
with a constant density are distributed over the fractal. 
In this case, the number of particles $N$ enclosed 
in a volume of characteristic size $R$ satisfies the 
scaling law 
\be  N(R) \sim R^{D} ,\ee 
whereas for a regular n-dimensional Euclidean object 
we have $N(R)\sim R^n$.

For distribution of particles with number density $n({\bf r},t)$,
we have that the mass density can be defined by
\be \rho({\bf r},t)=m n({\bf r},t) , \ee
where $m$ is the mass of a particle. 
The total mass of region $W$ is then given by the integral
\be M(W)=\int_W \rho({\bf r},t) dV_3 , \ee
or $M(W)=mN(W)$, where $N(W)$ is a number of particles in the region $W$. 
The fractional generalization of this equation can be written
in the following form: 
\be M(W)=\int_W \rho_D({\bf r},t) dV_D , \ee
where $D$ is a mass fractal dimension of the distribution,
and $dV_D$ is an element of D-dimensional volume such that
\be \label{5a} dV_D=C_3(D,{\bf r})dV_3. \ee
For the Riesz definition of the fractional integral \cite{SKM}, the 
function $C_3(D,{\bf r})$ is defined by the relation
\be \label{5R} C_3(D,{\bf r})=
\frac{2^{3-D}\Gamma(3/2)}{\Gamma(D/2)} |{\bf r}|^{D-3} . \ee
The initial points of the fractional integral are set to zero.
The numerical factor in Eq. (\ref{5R}) has this form in order to
derive usual integral in the limit $D\rightarrow (3-0)$.
Note that the usual numerical factor
$\gamma^{-1}_3(D)={\Gamma(1/2)}/{2^D \pi^{3/2} \Gamma(D/2)}$,
which is used in Ref. \cite{SKM}
leads to $\gamma^{-1}_3(3-0)= 1/[4 \pi^{3/2}]$ 
in the limit $D\rightarrow (3-0)$.

For the Riemann-Liouville fractional integral \cite{SKM}, 
the function $C_3(D,{\bf r})$ is defined by
\be \label{5RL} C_3(D,{\bf r})=
\frac{|x y z |^{D/3-1}}{\Gamma^3(D/3)}  . \ee
Here we use Cartesian's coordinates $x$, $y$, and $z$. 
In order to have the usual dimensions of the physical values,
we can use vector ${\bf r}$, and coordinates 
$x$, $y$, $z$ as dimensionless values.
Therefore the density $\rho_D$ has the dimension of mass.

Note that the interpretation of fractional integration
is connected with fractional dimension \cite{chaos,PRE05}.
This interpretation follows from the well-known formulas 
for dimensional regularizations \cite{Col}.
The fractional integral can be considered as an 
integral in the fractional dimension space
up to the numerical factor $\Gamma(D/2) /( 2 \pi^{D/2} \Gamma(D))$.

If we consider the ball region $W=\{{\bf r}: \ |{\bf r}|\le R \}$, 
and the spherically symmetric distribution of particles 
($n_D({\bf r},t)=n_D(r)$), then we have
\be N(R)=4\pi \frac{2^{3-D}\Gamma(3/2)}{\Gamma(D/2)}
\int^R_0 n_D(r) r^{D-1} dr , \quad M(R)=mN(R). \ee
For the homogeneous ($n_D(r)=n_0$) fractal distribution, we get
\be N(R)=4\pi n_0 \frac{2^{3-D}\Gamma(3/2)}{\Gamma(D/2)}
\frac{R^D}{D} \sim R^D . \ee
Fractal distribution of particles is called a homogeneous 
fractal distribution if the power law  $N(R)\sim R^D $ does not 
depend on the translation  of the region. 
The homogeneity property of the distribution
can be formulated in the following form:
for all regions $W$ and $W^{\prime}$ 
such that the volumes are equal $V(W)=V(W^{\prime})$, 
we have that the number of particles in these regions 
are equal too, $N(W)=N(W^{\prime})$. 
Note that the wide class of the fractal media satisfies 
the homogeneous property.
In many cases, we can consider the porous media \cite{Por1,Por2}, 
polymers \cite{P}, colloid aggregates \cite{CA}, and 
aerogels \cite{aero} as homogeneous fractal media.
In Refs. \cite{PLA05,AP05}, the continuous medium model for the fractal 
distribution was suggested. Note that the fractality and 
homogeneity properties can be realized 
for the fractional continuous model in the following forms: 

\noindent
(1) {\it Homogeneity}:
the local number density of homogeneous fractal distribution 
is translation invariant value that has the form
$n({\bf r})=n_0=const$.

\noindent
(2) {\it Fractality}:
the number of particles in the ball region $W$ obeys a power law relation
$N_D(W) \sim R^D$,
where $D<3$, and $R$ is the radius of the ball.

%%%%%%%%%%%%%%%%%%%%%%%%%%%%%%%%%%%%%%%%%%%%%%%%%%%%%%%%%%%%%%%%%%
\subsection{Flow of Fractal Medium}

For distribution of particles with number density $n({\bf r},t)$ flowing 
with velocity ${\bf u}={\bf u}({\bf r},t)$, 
the resulting density ${\bf J}({\bf r},t)$ is given by
\be {\bf J}({\bf r},t)= m n({\bf r},t) {\bf u} , \ee
where $m$ is the mass of a particle. 
Measuring the field ${\bf J}({\bf r},t)$ passing through a surface 
$S=\partial W$ gives the flow (flux of mass) 
\be I(S)=\Phi_J(S)=\int_S ({\bf J}, d{\bf S}_2) , \ee
where ${\bf J}={\bf J}({\bf r},t)$ is the flow field vector, 
$d{\bf S_2}=dS_2{\bf n}$ is a differential unit of area 
pointing perpendicular to the surface $S$, 
and the vector ${\bf n}=n_k{\bf e}_k$ is a vector of normal.
The fractional generalization of this equation for the fractal 
distribution can be written in the following form
\be I(S)=\int_S ({\bf J}({\bf r},t), d{\bf S}_d) , \ee
where we use
\be \label{C2} dS_d=C_2 (d,{\bf r})dS_2 , \quad 
C_2(d,{\bf r})= \frac{2^{2-d}}{\Gamma(d/2)} |{\bf r}|^{d-2} . \ee
Note that $C_2(2,{\bf r})=1$ for $d=2$. 
In general, the medium on the boundary $\partial W$ has the dimension $d$. 
In the general case, the dimension $d$ is not equal to $2$ and 
is not equal to $(D-1)$.

\subsection{Equation of Continuity for Fractal Distribution}

The change of mass inside a region $W$ bounded 
by the surface $S=\partial W$ is always equal 
to the flux of mass through this surface.
This is known as the law of mass conservation or the 
equation of balance of mass \cite{AP05}. If we denote
by ${\bf J}({\bf r},t)$ the flow density, then mass conservation is written
\be \frac{dM(W)}{dt}=-I(S), \ee
or, in the form
\be \label{cecl} \frac{d}{dt} \int_W \rho_D({\bf r},t) dV_D= 
- \oint_{\partial W} ({\bf J} ({\bf r},t),d{\bf S}_d) . \ee
In particular, when the surface $S=\partial W$ is fixed,
we can write
\be \label{drho} \frac{d}{dt} \int_W \rho_D({\bf r},t) dV_D= 
\int_W \frac{\partial \rho_D({\bf r},t)}{\partial t} dV_D .\ee
Using the fractional generalization of the 
mathematical Gauss's theorem (see Appendix), we have
\be \label{gt} \oint_{\partial W} ({\bf J} ({\bf r},t),d{\bf S}_d) =
\int_W C^{-1}_3(D,{\bf r})
\frac{\partial}{\partial x_k} \Bigl( C_2(d,{\bf r})J_k({\bf r},t) \Bigr)
dV_D .\ee
Substituting the right hand sides of Eqs. (\ref{drho}) and (\ref{gt})
in Eq. (\ref{cecl}), we find the equation of balance of mass 
in differential form
\be C_3(D,{\bf r})\frac{\partial \rho_D({\bf r},t)}{\partial t}+
\frac{\partial}{\partial x_k} \Bigl( C_2(d,{\bf r})J_k({\bf r},t) \Bigr) =0 . \ee
This equation can be considered as a continuity equation for
fractal distribution of particles \cite{AP05}.

\section{Gravitational Field of Fractal Distribution}

\subsection{Gravitational Field}

For a point mass $M$ at position  ${\bf r}^{\prime}$ 
the gravitational field ${\bf F}$ at a point ${\bf r}$ is defined  by
\be {\bf F}=G M \
\frac{{\bf r}^{\prime}-{\bf r}}{|{\bf r}^{\prime}-{\bf r}|^3} , \ee
where $G$ is a gravitational constant.  
 
For a continuous distribution $\rho({\bf r}^{\prime})$ of mass, 
the gravitational field ${\bf F}$ at a point ${\bf r}$ is given  by
\be \label{E}
{\bf F}_3({\bf r})=G \int_W
\frac{{\bf r}^{\prime}-{\bf r}}{|{\bf r}^{\prime}-{\bf r}|^3}
\rho({\bf r}^{\prime}) dV^{\prime}_3 . \ee
The fractional generalization of this equation 
for a fractal distribution of mass is given by the equation
\be \label{CLD}
{\bf F}_D({\bf r})=G \int_W
\frac{{\bf r}^{\prime}-{\bf r}}{|{\bf r}^{\prime}-{\bf r}|^3}
\rho_D({\bf r}^{\prime}) dV^{\prime}_D , \ee
where $dV^{\prime}_D=C_3(D,{\bf r}^{\prime}) dV^{\prime}_3$. 
Eq. (\ref{CLD}) can be considered as Newton's law written 
for a fractal distribution of particles. 

Measuring the gravitational field passing through a surface 
$S=\partial W$ gives the gravitational filed flux 
\be \Phi_F(S)=\int_S ({\bf F}, d{\bf S}_2) , \ee
where ${\bf F}$ is the gravitational field vector, and $d{\bf S}_2$ 
is a differential unit of area pointing perpendicular to the surface $S$.

\subsection{Gauss's Law for Fractal Distribution}

The Gauss's law tells us that the total flux $\Phi_F(S)$ of 
the gravitational field ${\bf F}$ through a closed surface $S=\partial W$ 
is proportional to the total mass $M(W)$ inside the surface: 
\be \label{GL1} \Phi_F(\partial W)=4 \pi G M(W) . \ee
For the fractal distribution, Gauss's law states
\be \label{GL2} \int_S ({\bf F}_D,d{\bf S}_2)=4 \pi G 
\int_W \rho_D ({\bf r}) dV_D , \ee 
where ${\bf F}={\bf F}({\bf r})$ is the gravitational field, and 
$\rho_D({\bf r})$ is the mass density,  $dV_D=C_3(D,{\bf r})dV_3$, 
and $G$ is the gravitational constant. 

Gauss's law  by itself can be used to find the gravitational field 
of a point mass at rest, and the principle of superposition 
can then be used to find the gravitational field of an arbitrary 
fractal mass distribution.

If we consider the spherically symmetric fractal distribution
$\rho_D({\bf r})=\rho_D(r)$, and the ball region 
$W=\{{\bf r}:\ |{\bf r}|\le R\}$, then we have
\be M(W)=4 \pi \int^R_0 \rho_D(r) C_3(D,{\bf r}) r^2 dr , \ee
where $C_3(D,{\bf r})$ is defined in Eq. (\ref{5R}), i.e., 
\be \label{MW} M(W)=4 \pi \frac{2^{3-D}\Gamma(3/2)}{\Gamma(D/2)}
\int^R_0 \rho_D(r) r^{D-1} dr . \ee
Using the sphere $S=\{{\bf r}: \ |{\bf r}|= R \}$ as a 
surface $S=\partial W$, we get
\be \label{PW} \Phi_F(\partial W)= 4 \pi R^2 F_D(R). \ee
Substituting Eqs. (\ref{MW}) and (\ref{PW}) in the Gauss's law (\ref{GL1}),
we get the equation for gravitational field.
As the result, the Gauss's law for fractal distribution 
with spherical symmetry leads us to the equation for gravitational field
\be F_D(R)=\frac{ \pi G 2^{5-D}\Gamma(3/2)}{R^2 \Gamma(D/2)}
\int^R_0 \rho_D(r) r^{D-1} dr .\ee
For example, the gravitational field of homogeneous ($\rho_D(\bf r)=\rho_0$)
spherically symmetric fractal distribution of mass is defined by
\be F_D(R)=\rho_0 \frac{ \pi G 2^{5-D}\Gamma(3/2)}{ D \Gamma(D/2)} 
R^{D-2}  \sim R^{D-2} .\ee

%%%%%%%%%%%%%%%%%%%%%%%%%%%%%%%%%%%%%%%%%%%%%%%%%%%%%%%%%%%%%
\section{Multipole Moments for Fractal Distribution}

\subsection{Multipole Expansion}

A multipole expansion is a series expansion of the effect produced 
by a given system in terms of an expansion parameter which becomes 
small as the distance away from the system increases. 
Therefore, the leading one of the terms in a multipole expansion are 
generally the strongest. The first-order behavior of the system 
at large distances can therefore be obtained from the first terms 
of this series, which is generally much easier to compute than 
the general solution. Multipole expansions are most commonly used 
in problems involving the gravitational field of mass aggregations, 
the gravity and magnetic fields of mass and flow distributions, 
and the propagation of electromagnetic waves.

To compute one particular case of a multipole expansion, 
let ${\bf R}=X_k {\bf e}_k$ be the vector from a fixed reference point 
to the observation point; 
${\bf r}=x_k {\bf e}_k$ be the vector from the reference 
point to a point in the distribution; and ${\bf d}={\bf R}-{\bf r}$
be the vector from a point in the distribution to the observation point. 
The law of cosines then yields
\be 
d^2=R^2\Bigl( 1+\frac{r^2}{R^2} -2\frac{r}{R} \cos \; \theta \Bigr) , \ee
where $d=|{\bf d}|$, and $\cos \; \theta = ({\bf r},{\bf R})/(r R)$, so
\be d=R \sqrt{ 1+\frac{r^2}{R^2} -2\frac{r}{R} \cos \; \theta } . \ee
Now define $\epsilon ={r}/{R}$, and $x=\cos \; \theta$, then
\be \frac{1}{d}=\frac{1}{R} 
\Bigl( 1-2 \epsilon x+\epsilon^2 \Bigr)^{-1/2}. \ee
But  $\Bigl( 1-2 \epsilon x+\epsilon^2 \Bigr)^{-1/2}$
is the generating function for Legendre polynomials $P_n(x)$ as follows: 
\be \Bigl( 1-2 \epsilon x+\epsilon^2 \Bigr)^{-1/2}=
\sum^{\infty}_{n=0} \epsilon^n P_n(x) , \ee
so, we have the equation
\be \frac{1}{d}=\frac{1}{R} \sum^{\infty}_{n=0} 
\Bigl(\frac{r}{R}\Bigr)^n P_n( \cos \; \theta) . \ee
The gravitational potential $U$ (${\bf F}=-\nabla U$)
obeys ($1/d$) law.  Therefore, this potential
can be expressed as a multipole expansion
\be \label{11}
U=  -G  \sum^{\infty}_{n=0} \frac{1}{R^{n+1}}
\int_W r^n P_n( \cos \; \theta)  \rho_D({\bf r}) dV_D. 
\ee
The $n = 0$ term of this expansion 
can be pulled out by noting that $P_0(x)=1$, so
\be
U= - \frac{G}{R} \int_W \rho_D({\bf r}) dV_D-
G \sum^{\infty}_{n=1} \frac{1}{R^{n+1}}
\int_W r^n P_n( \cos \; \theta)  \rho_D({\bf r}) dV_D. 
\ee
The nth term
\be
U_n=- \frac{G}{R^{n+1}}\int_W r^n P_n( \cos \; \theta) \rho_D({\bf r}) dV_D 
\ee
is commonly named multipole, and for $n=2$ -  quadrupole.

\subsection{Gravitational Moment of Fractal Distribution}

Gravitational moments describe the nonuniform distribution of mass. 
The gravitational quadrupole term is given by 
\be U_2=- \frac{G}{R^3} \int_W r^2 P_2(\cos \; \theta) 
\rho_D({\bf r}) dV_D =\ee
\be =-\frac{G}{2 R^3} \int_W r^2 (3 \cos^2 \; \theta-1) 
\rho_D({\bf r}) dV_D =\ee
\be =-\frac{G}{2 R^3} \int_W 
( \frac{3}{R^2}({\bf R},{\bf r})^2-r^2) \rho_D({\bf r}) dV_D .\ee
The quadrupole is the third term in a gravitational 
multipole expansion, and can be defined by
\be U_2= - \frac{G}{2 R^3} \sum^3_{k,l=1} \frac{X_k X_l}{R^2} M_{kl} , \ee
where $G$ is the gravitational constant, 
$R$ is the distance from the fractal distribution of mass, and 
$M_{kl}$  is the gravitational quadrupole moment, which is a tensor.

The gravitational quadrupole moment is defined by the equation
\be M_{kl}=\int_W (3 x_k x_l-r^2\delta_{kl}) \rho_D({\bf r}) dV_D ,\ee
where $x_k= x, y$, or $z$. From this definition, it follows that
\be M_{kl}=M_{lk} , \quad and  \quad \sum^{3}_{k=1} M_{kk}=0. \ee
Therefore, we have $M_{zz}=-M_{xx}-M_{yy}$.
In order to compute the values
\be M^{(D)}_{xx}=
\int_W  [2x^2-y^2-z^2] \rho_D({\bf r}) dV_D , \ee 
\be M^{(D)}_{yy}=
\int_W [-x^2+2y^2-z^2)] \rho_D({\bf r}) dV_D , \ee 
\be M^{(D)}_{zz}=
\int_W [-x^2-y^2+2z^2)] \rho_D({\bf r}) dV_D , \ee 
we consider the following expression
\be \label{Mabc} M(\alpha,\beta,\gamma)=
\int_W [\alpha x^2+\beta y^2+\gamma z^2)] \rho_D({\bf r}) dV_D ,\ee
where we use the Riemann-Liouville fractional integral \cite{SKM}, and  
the function $C_3(D,{\bf r})$ in the form
\be C_3(D,{\bf r})=
\frac{|x y z |^{a-1}}{\Gamma^3(a)} , \quad a=D/3. \ee
Using Eq. (\ref{Mabc}), we have
\be \label{MM} M^{(D)}_{xx}=M(2,-1,-1), \quad  M^{(D)}_{xx}=M(-1,2,-1),
\quad M^{(D)}_{zz}=M(-1,-1,2) . \ee

\subsection{Quadrupole Moment of Fractal Parallelepiped}

Let us consider the example of gravitational quadrupole moment
for the homogeneous ($\rho_D({\bf r})=\rho_0$) fractal distribution
in the parallelepiped region
\be\label{par} 
W=\{(x;y;z): \ 0 \le x \le A,\  0 \le y \le B , \ 0 \le z \le C \} . \ee
If we consider the region $W$ in the form (\ref{par}), then we get
(\ref{Mabc}) in the form
\be M(\alpha,\beta,\gamma)= \frac{\rho_0 (ABC)^a}{(a+2)a^2 \Gamma^3(a) }
(\alpha A^2+\beta B^2+\gamma C^2) . \ee
The total mass of this region $W$ is 
\be M(W)=\rho_0 \int_W dV_D=\frac{\rho_0 (ABC)^a}{a^3 \Gamma^3(a)}  .\ee
Therefore, we have the following equation
\be M(\alpha,\beta,\gamma)= \frac{a}{a+2} M(W)
(\alpha A^2+\beta B^2+\gamma C^2) , \ee
where $a=D/3$. If $D=3$, then we have
\be M(\alpha,\beta,\gamma)= \frac{1}{3} M(W)
(\alpha A^2+\beta B^2+\gamma C^2) . \ee
As the result, we get gravitational quadrupole moments $M^{(D)}_{kk}$ of 
fractal distribution in the region $W$:
\be M^{(D)}_{kk}=\frac{3D}{D+6} \ M^{(3)}_{kk} , \ee
where $M^{(3)}_{kk}$ are moments for the usual 
homogeneous distribution ($D=3$). 
By analogy with these equations, we can derive $M^{(D)}_{kl}$ for the case
$k\not=l$. These quadrupole moments are
\be M^{(D)}_{kl}=\frac{4 D^2}{(D+3)^2} \ M^{(3)}_{kl} , \quad (k\not=l). \ee
Using $2<D\le 3$, we get the relations 
\be 0.75 < \frac{3D}{D+6}\le 1 , \quad
0.64 < \frac{4 D^2}{(D+3)^2} \le 1 . \ee
Quadrupole moment of fractal ellipsoid is considered in Appendix.

\section{Conclusion}

The fractional continuous models for fractal distribution of particles
can have a wide application. 
This is due in part to the relatively small numbers of parameters 
that define a random fractal distribution of great complexity
and rich structure.
In many cases, the real fractal structure of matter can be disregarded 
and the distribution of particles can be replaced by  
some fractional continuous model \cite{PLA05,AP05}. 
In order to describe the distribution with 
non-integer dimension, we must use the fractional calculus.
Smoothing of the microscopic characteristics over the 
physically infinitesimal volume transforms the initial 
fractal distribution into fractional continuous model
that uses the fractional integrals. 
The order of fractional integral is equal 
to the fractal dimension of the distribution.
The fractional continuous model for the fractal distribution
allows us to describe dynamics 
of a wide class of fractal media \cite{AP05,Chaos05,Physica}.  

The suggested results can have a wide application 
to galactic dynamics and cosmology. 
In particular, there is strong evidence that the distribution of 
mass beyond the scale of clusters of galaxies is fractal, 
with $D \simeq 1.2$, corresponding to a power-law two-point 
correlation function with exponent equal to $-1.8$. 
Fractal distributions may also be present within gravitational 
systems of a smaller scale, for example, galaxies.
However, the results are incomplete in the following aspect: 
it is known that the fractal distribution of mass in 
the Universe is characterized by large density fluctuations, 
even within the fractal volume, therefore it is far from homogeneous. 
The nonhomogeneity of the fractal distribution can be 
described by the suggested fractional continuous model \cite{PLA05}.
The fluctuation deviation from homogeneity can be parametrized 
by the so-called n-point correlation functions, with $n>2$ 
\cite{N1,N2,N3,N4}. 
Such density fluctuations are important in that they produce 
terms in the force field which can be described only statistically. 
An elementary example is the random distribution of particles 
in a three-dimensional sphere. 
Although the distribution can be considered as uniform 
when viewed at the scale of the sphere (with $D=3$), 
the Poisson noise of the density field will cause gravitational 
clustering in small scales that will finally prevail 
the overall evolution of the system, i.e., 
the latter will be quite a different from the evolution of 
a perfectly homogeneous sphere.

\section*{Appendix}

\subsection*{Fractional Gauss's theorem}

In order to realize the representation, we derive the fractional 
generalization of the Gauss's theorem
\be \label{GT}
\int_{\partial W} ({\bf J}({\bf r},t), d{\bf S}_2) 
=\int_W div( {\bf J}({\bf r},t) ) dV_3 ,
\ee
where the vector ${\bf J}({\bf r},t)=J_k{\bf e}_k$ is a field, and
\be div( {\bf J})=\frac{\partial {\bf J}}{\partial {\bf r}}= 
\frac{\partial J_k}{\partial x_k} . \ee
Here and later we mean the sum on the repeated index $k$ from 1 to 3.
Using Eq. (\ref{C2}), we get
\be \int_{\partial W} ({\bf J}({\bf r},t),d{\bf S}_d) 
=\int_{\partial W}  C_2(d,{\bf r})  ({\bf J}({\bf r},t) , d{\bf S}_2) . \ee
Note that we have $C_2(2,{\bf r})=1$ for the $d=2$. 
Using the usual Gauss's theorem (\ref{GT}), we get 
\be \int_{\partial W}  C_2(d,{\bf r}) ({\bf J}({\bf r},t), d{\bf S}_2) =
\int_W  div(C_2(d,{\bf r}) {\bf J}({\bf r},t)) dV_3 . \ee
Equations (\ref{5a}) and (\ref{5R}) in the form 
$dV_3=C^{-1}_3(D,{\bf r}) dV_D$
allows us to derive the fractional generalization of the Gauss's theorem:
\be \label{FGT}
\int_{\partial W} ({\bf J}({\bf r},t), d{\bf S}_d)=
\int_W C^{-1}_3(D,{\bf r}) 
div \Bigr( C_2(d,{\bf r}) {\bf J}({\bf r},t) \Bigr) \ dV_D .
\ee

\subsection*{Quadrupole Moment of Fractal Ellipsoid}

Let us consider the example of gravitational quadrupole moment
for the homogeneous ($\rho_D({\bf r})=\rho_0$) fractal distribution
in the ellipsoid region $W$:
\be\label{ell} 
\frac{x^2}{A^2}+\frac{y^2}{B^2}+\frac{z^2}{C^2} \le 1 . \ee
If we consider the region $W$ in the form (\ref{ell}), then we get
(\ref{Mabc}) in the form
\be M(\alpha,\beta,\gamma)= \frac{\rho_0 (ABC)^a}{ (3a+2) \Gamma^3(a) }
(\alpha A^2 K_1(a)+\beta B^2 K_2(a)+\gamma C^2 K_3 (a) ) , \ee
where $a=D/3$, and $K_i(a)$ (i=1,2,3) are defined by
\be K_1(a)=L(a+1,a-1,2\pi) L(a-1,2a+1,\pi), \ee
\be K_2(a)=L(a-1,a+1,2\pi) L(a-1,2a+1,\pi),  \ee
\be K_3(a)=L(a-1,a-1,2\pi) L(a+1,2a-1,\pi) . \ee
Here we use the following function
\be L(n,m,l)=\frac{2l}{\pi} \int^{\pi/2}_0 
dx \ |\cos (x)|^{n} | \sin (x) |^m =
\frac{l}{\pi} \frac{\Gamma(n/2+1/2) \Gamma(m/2+1/2)}{\Gamma(n/2+m/2+1)}. 
\ee
If $D=3$, we obtain
\be \label{Mabce} 
M(\alpha,\beta,\gamma)= \frac{4\pi}{3} \frac{\rho_0 ABC}{5 }
(\alpha A^2 +\beta B^2 +\gamma C^2 ) , \ee
where we use $K_1=K_2=K_3={4\pi}/{3}$.
The total mass of this region $W$ is 
\be \label{Me} M(W)=\rho_0 \int_W dV_D=\frac{\rho_0 (ABC)^a}{3 a \Gamma^3(a)} 
\frac{2 \Gamma^3(a/2)}{\Gamma(3a/2)} .\ee
If $D=3$, we have the total mass
\be M(W)=\rho_0 \int_W dV_3=\frac{4 \pi}{3} \rho_0 ABC . \ee

Using Eqs. (\ref{Mabce}) and (\ref{Me}), we get the 
quadrupole moments (\ref{MM}) for fractal ellipsoid
\be M(\alpha,\beta,\gamma)= \frac{a M(W)}{3a+2} 
(\alpha A^2 +\beta B^2 +\gamma C^2 ) , \ee
where $a=D/3$. If $D=3$, then we have the well-known relation
\be M(\alpha,\beta,\gamma)= \frac{M(W)}{5}
(\alpha A^2+\beta B^2+\gamma C^2) . \ee

%%%%%%%%%%%%%%%%%%%%%%%%%%%%%%%%%%%%%%%%%%%%%%%%%%%%%%%%%%%

%%%%%%%%%%%%%%%%%%%%%%%%%%%%%%%%%%%%%%%%%%%%%%
\end{document}